# Ultrathin All-in-one Spin Hall Magnetic Sensor with Built-in AC Excitation Enabled by Spin Current

*Yanjun Xu, Yumeng Yang, Mengzhen Zhang, Ziyan Luo and Yihong Wu\**

Department of Electrical and Computer Engineering, National University of Singapore
4 Engineering Drive 3, Singapore 117583
*Email: elewuyh@nus.edu.sg

Keywords: spin current, spin-orbit torque, spin-Hall magnetoresistance sensor, AC excitation

**Magnetoresistance (MR) sensors provide cost-effective solutions for diverse industrial and consumer applications, including emerging fields such as internet-of-things (IoT), artificial intelligence and smart living. Commercially available MR sensors such as anisotropic magnetoresistance (AMR) sensor, giant magnetoresistance (GMR) sensor and tunnel magnetoresistance (TMR) sensors typically require an appropriate magnetic bias for both output linearization and noise suppression, resulting in increased structural complexity and manufacturing cost. Here, we demonstrate an all-in-one spin Hall magnetoresistance (SMR) sensor with built-in AC excitation and rectification detection, which effectively eliminates the requirements of any linearization and domain stabilization mechanisms separate from the active sensing layer. This was made possible by the coexistence of SMR and spin-orbit torque (SOT) in ultrathin NiFe/Pt bilayers. Despite the simplest possible structure, the fabricated Wheatstone bridge sensor exhibits essentially zero DC offset, negligible hysteresis, and a detectivity of around $1 nT/\sqrt{Hz}$ at 1Hz. In addition, it also shows an angle dependence to external field similar to those of GMR and TMR, though it does have any reference layer (unlike GMR and TMR). The superior performances of SMR sensors are evidently demonstrated in the proof-of-concept experiments on rotation angle measurement, and vibration and finger motion detection.**

Magnetic field sensing is so important that each time when a new magnetic or spintronic phenomenon was discovered there would be an attempt to exploit it for magnetic sensing applications with improved cost-performances. The most notable examples in recent years are the giant magnetoresistance (GMR) and tunnel magnetoresistance (TMR) in ultrathin magnetic/non-magnetic heterostructures.[1] Together with anisotropic magnetoresistance (AMR), these magnetoresistance (MR) effects have led to a wide range of compact and high-sensitivity magnetic sensors for diverse industrial and consumer applications;[2] these sensors are expected to play even more important roles in the rapidly developing internet-of-things (IoT) paradigm and related technologies,[3] which require 100 trillion of sensors by 2030. The latest additions to these fascinating spintronic effects are spin orbit torque (SOT)[4] and spin Hall magnetoresistance (SMR)[5] in ferromagnet (FM) / heavy metal (HM) bilayers. Taking advantage of these intriguing effects, recently we have demonstrated an AMR/SMR sensor (hereafter we call it SMR sensor considering the fact that SMR is dominant) with the SOT effective field as the built-in linearization mechanism,[6,7] which effectively replaces the sophisticated linearization mechanism employed in conventional MR sensors.[8] However, as the sensors were driven by DC current, we still faced the same issues as commercial AMR sensors, that is, DC offset and domain motion induced noise. Here, we demonstrate that, by introducing AC excitation, we achieved an all-in-one magnetic sensor which embodies multiple functions of AC excitation, domain stabilization, rectification detection, and DC offset cancellation, and importantly, all these features are realized in a simplest possible structure which consists of only an ultrathin NiFe/Pt bilayer. Such kind of integrated AC excitation and rectification are not possible in conventional AMR sensors. The sensors are essentially free of DC offset with negligible hysteresis and low noise (with a detectivity of 1nT/$\sqrt{Hz}$ at 1Hz). Through a few proof-of-concept experiments, we show that these sensors promise great potential in a variety of low-field sensing applications including navigation, angle detection, and wearable electronics.

Both SOT and SMR appear when a charge current passes through a FM/HM bilayer. Although the exact mechanism still remains debatable, both Rashba[9] and spin Hall effect (SHE)[10] are commonly believed to play a crucial role in giving rise to the SOT in FM/HM bilayers. There are two types of SOTs, one is field-like (FL) and the other is damping-like (DL); the latter is similar to spin transfer torque. Phenomenologically, the two types of torques can be modelled by $\vec{T}_{DL} = \tau_{DL}\vec{m} \times [\vec{m} \times (\vec{j} \times \vec{z})]$ and $\vec{T}_{FL} = \tau_{FL}\vec{m} \times (\vec{j} \times \vec{z})$, respectively, where $\vec{m}$ is the magnetization direction, $\vec{j}$ is the in-plane current density, $\vec{z}$ is the interface normal, and $\tau_{FL}$ and $\tau_{DL}$ are the magnitude of the FL and DL torques, respectively.[11] Corresponding to the two torques are two effective fields, one is damping-like ($H_{DL}$) and the other is field-like ($H_{FL}$). On the other hand, the SMR generated in the bilayer is given by $-\Delta R_{SMR}(\vec{m} \cdot \vec{\sigma})^2$, where $\Delta R_{SMR}$ is the change in resistance induced by the SMR effect, and $\vec{\sigma}$ is the polarization direction of the spin current. When the magnetization rotates in the plane (which is of interest in this work), the total longitudinal resistance is given by $R_{xx} = R_o - (\Delta R_{SMR} + \Delta R_{AMR})(\vec{m} \cdot \vec{\sigma})^2$, where $R_o$ is the longitudinal resistance and $\Delta R_{AMR}$ is the resistance change caused by AMR. For ultrathin FM/HM bilayer, $\Delta R_{SMR}$ is typically 2-3 times larger than $\Delta R_{AMR}$.[12]

We now consider a Wheatstone bridge with four elliptically shaped elements made of NiFe/Pt bilayer. The dimension, $a \times b \times t$, is the same for all the four elements, with $a$ ($b$) and $t$ the length of long-axis (short-axis) and thickness, respectively. The easy axis is in the long-axis or $x$-direction, whereas the hard axis is in the short-axis or $y$-direction. When an AC current, $I = I_o sin\omega t$, is applied to two terminals of the bridge, the voltage across the other two terminals is given by (see Supporting Information Section S1)

$$V_{out} = \frac{1}{2}I_o \Delta R_0 sin\omega t + \frac{1}{2}\frac{\alpha I_0^2 \Delta R H_y cos2\omega t}{(H_D+H_K)^2} - \frac{1}{2}\frac{\alpha I_0^2 \Delta R H_y}{(H_D+H_K)^2} \quad (1)$$

where $\Delta R_o$ is the offset resistance between the neighboring sensing elements, $\Delta R = \Delta R_{SMR} + \Delta R_{AMR}$, $H_D$ is the demagnetizing field, $H_K$ is the uniaxial anisotropy filed, $H_y$ is the externally applied magnetic field or field to be detected in *y*-direction, and $\alpha$ is the ratio between the current induced bias field $H_{bias}$, which is the sum of field-like SOT effective field $H_{FL}$ and the Oersted field, and the applied current, *i.e.*, $H_{bias} = H_{FL} + H_{Oe} = \alpha I$. Although the Oersted field ($H_{Oe}$) from the current is also in the same direction as $H_{FL}$ in NiFe/Pt bilayer structure, its magnitude is generally much smaller compared to $H_{FL}$.[7] The time-average or DC component of $V_{out}$ is given by

$$\overline{V_{out}} = \frac{\alpha I_0^2 \Delta R}{2(H_D+H_K)^2} H_y \qquad (2)$$

We can see that, by simply replacing the DC current with an AC current, we obtained two significant results: linear response to external field ($H_y$) and zero DC offset. It is worth noting that, under the AC excitation, it is no longer necessary to bias the magnetization 45° away from the easy axis for output linearization, which greatly simplifies the sensor design. The DC offset, if any, caused by process fluctuations, can be effectively eliminated by this technique. It is obvious from Equation (1) that, in addition to DC detection, the same output voltage can also be extracted from the 2nd harmonic using a standard lock-in technique. In addition to intrinsic linearity and zero DC offset, the AC excitation also effectively suppresses the hysteresis and noise. As we will discuss shortly in experimental section, we are able to completely eliminate the hysteresis and significantly reduce the noise by using the AC excitation technique. It is worth pointing out that such kind of built-in AC excitation cannot be implemented in conventional AMR, GMR or TMR sensors.

**Figure 1**a shows the scanning electron micrograph (SEM) of the fabricated Wheatstone bridge sensor comprising of four ellipsoidal NiFe(1.8 nm)/Pt(2 nm) bilayer sensing elements with a long axis length (*a*) of 800 μm and an aspect ratio of 4:1 (see Experimental Section). The thicknesses have been optimized to give the largest SOT and SMR.[6,7] The spacing (*L*)

between the two electrodes for each element is kept at *a*/3. As shown in the schematic of Figure 1b, the sensor is driven by an AC current and its response to an external magnetic field ($H_y$) is detected as a DC voltage. In order to minimize the influence of earth field, both the sensor and Helmholtz coils for generating $H_y$ were placed inside a magnetically shielded cylinder made of 7 layers of µ-metals. The AC current serves as both the source for an AC excitation field and the sensing current. Figure 1c shows the response of the sensor to an external field swept in a full loop, *i.e.*, from -0.5 Oe to +0.5 Oe and then back to -0.5 Oe in *y*-direction. The root mean square (rms) amplitude and frequency of the applied AC bias current density are $5.5 \times 10^5$ A/cm$^2$ and 5000 Hz, respectively. The current density was chosen such that a linear response with maximum sensitivity is achieved near zero field, though frequency is less critical (Supporting Information Section S2). Since ultrathin NiFe is very soft, the dynamic range of the sensor is essentially determined by the shape anisotropy.[6,7] We chose a field range such that the magnetic bias can be readily achieved using the SOT effective field with a reasonably small bias current, which is about 0.8 Oe at a current density of $10^6$ A/cm$^2$ for NiFe (1.8 nm) / Pt (2 nm) bilayer.[7] It is apparent from Figure 1c that the response curves for forward and backward sweeping almost completely overlap with each other, indicating a negligible hysteresis even in a full field range sweeping. Another important characteristic to note is that the DC offset is nearly zero, as expected from Equation (2). As shown in the inset, with fine tuning of the bias current, the offset can be suppressed to nearly zero (below 1 µV in this case and is limited by the electronics). The sensitivity extracted from the response curve is about 1.17 mV/V/Oe, which is comparable to the sensitivity of commercial AMR sensors (0.8 ~ 1.2 mV/V/Oe) despite its ultrathin thickness.

Thermal stability is of great importance for practical applications. To evaluate temperature sensitivity of SMR sensor, we measured the temperature dependence of SMR and AMR for one of the sensing elements of the bridge sensor, and the results are shown in Figure 1d. From the figure, we can observe that SMR ratio is much less sensitive to temperature than AMR; the latter drops about 66% from 250 K to 400 K, whereas the former, *i.e.*, SMR, remains almost constant. For NiFe (1.8 nm)/Pt (2 nm) bilayer, the SMR is about two times larger than that of AMR, or in other words, around 2/3 of the MR signal comes from the SMR, and 1/3 is from AMR at room temperature. Therefore, the SMR sensor is less sensitive to thermal effect compared with conventional AMR sensors. Besides environmental temperature fluctuations, heating due to the bias current may also affect the stability of the sensor. In order to examine if there is any heating effect, we performed AC field sensing experiment for the same sensor whose quasi-static field response is shown in Figure 1c for a duration of 3 hours. During the measurement, an sinusoidal AC magnetic field with an amplitude of 0.1 Oe and a frequency of 0.1 Hz was applied in *y*-direction, while the sensor was biased by an AC current with a rms density of $5.5 \times 10^5$ A/cm$^2$ and a frequency of 5000 Hz. The sensor output voltage was monitored continuously for the entire duration. Figure 1e shows the sampled waveforms at zeroth, 1$^{st}$, 2$^{nd}$, and 3$^{rd}$ hour for a duration of 50s. As can be seen, there were no visible changes in both the amplitude and DC offset throughout the 3hrs duration. In order to analyze the sensor's stability more quantitatively, the average amplitude and offset of every 70 cycles were extracted from the raw data and their changes with respect to the initial values normalized by the signal amplitude are plotted in Figure 1f as a function of time. It turned out that both changes are very small; the signal amplitude changes by 0.15% throughout the 3hrs measurement period, while the offset varies about 0.25%. For comparison, we performed the same measurements for commercial HMC1001 AMR sensor (note: for a fair comparison, the measurements for both sensors were conducted without additional offset compensation). The signal amplitude change within the same duration is about 0.4%, but the initial DC offset is 83% which fluctuates around 3% throughout the 3hrs measurement duration (see Supporting Information Section S4 for more details). These results demonstrate clearly the excellent thermal stability of AC biased SMR sensors.

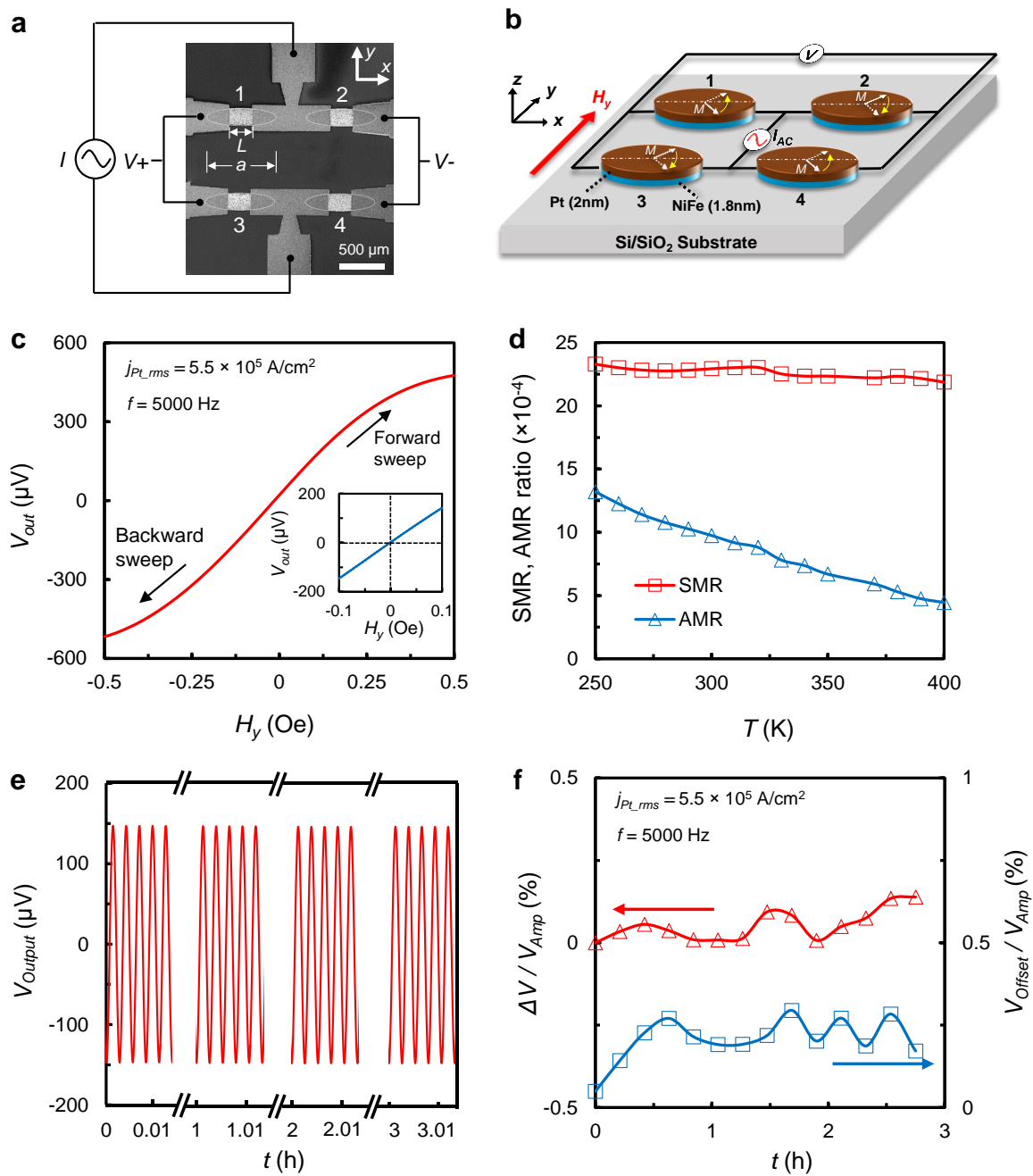

**Figure 1.** Field sensing performance of SMR sensor under AC excitation and DC detection. a) SEM image and schematic of the AC biased Wheatstone bridge SMR sensor. Scale bar: 500 μm. b) Schematic of a Wheatstone bridge SMR sensor comprised of four ellipsoidal NiFe/Pt bilayer sensing elements with the arrows indicating the magnetization direction driven by the AC current. c) Voltage output of the sensor in response to an external field swept in a full loop, *i.e.*, from -0.5 Oe to +0.5 Oe and then back to -0.5 Oe in *y*-direction. The rms amplitude and frequency of the applied AC bias current density are $5.5 \times 10^5$ A/cm$^2$ and 5000 Hz, respectively. Inset shows the response curve in a smaller sweeping range, which shows nearly a zero DC offset. d) Temperature dependence of AMR and SMR ratio for one of the sensing elements. e) Sampled waveforms of voltage output at zeroth, 1st, 2nd, and 3rd hour for a duration of 50s, in response to a sinusoidal AC magnetic field with an amplitude of 0.1 Oe and a frequency of 0.1 Hz applied in *y*-direction; the data are extracted from the measured waveform for the entire duration of 3 hours. During the measurement, the sensor was biased by an AC current with a rms density of $5.5 \times 10^5$ A/cm$^2$ and a frequency of 5000 Hz. f) Extracted changes of average amplitude and offset of every 70 cycles normalized by the initial value of signal amplitude as a function of time.

We now turn to the noise characteristics and detectivity of the sensor biased by both a DC and AC current (see Supporting Information Section S5 for the measurement setup). **Figure 2**a shows the detectivity of the sensor (the same sensor in Figure 1) under DC and AC bias at different frequencies. The root-mean-square current density of the AC bias was fixed at $5.5 \times 10^5$ A/cm$^2$ for different frequencies and is the same as the DC current density. The DC biased sensor exhibits a detectivity of about 2.8 nT/√Hz at 1 Hz. In comparison, the detectivity for AC biased sensor is about 1 nT/√Hz at 1 Hz, at all bias frequencies from 500 Hz to 100 kHz. The detectivity for AC biased sensor is constantly smaller from 0.1 Hz to 1000 Hz when the bias frequency is above 5 kHz. The increased noise level above 50 Hz for bias frequency at 500 Hz is caused by inadequate filtering of the AC signal from DC output in the measurement setup, as manifested by the large peak at 500 Hz. These noises can be readily reduced by employing a properly designed low-noise filter.

The noise at low frequency is dominated by 1/*f* noise, which in the case of magnetic sensors, generally consists of a non-magnetic[13] and a magnetic component.[14] The former originates from resistance fluctuations, while the latter is commonly attributed to thermally induced domain wall nucleation and motion[15] which can be described in terms of the fluctuation-dissipation (FD) relation for magnetization.[16] The power density of 1/*f* noise in the Wheatstone bridge SMR sensor can be described phenomelogically as:[17,18]

$$S_V = \frac{\delta_H V_b^2}{N_c \cdot Vol \cdot f} \quad (3)$$

where $\delta_H$ is the Hooge constant, $V_b$ is the bias voltage across the bridge, $N_c$ is the free electron density, *Vol* is the effective volume of NiFe and *f* is the frequency. The Hooge constant $\delta_H$ is a parameter characterizing the amplitude of the 1/*f* noise fluctuations. We first estimated the non-magnetic contribution by saturating the magnetization in easy axis direction and measuring the noise spectrum, from which a $\delta_H$ value of $1.3 \times 10^{-3}$ is obtained by using $N_c = 1.7 \times 10^{29}$ m$^{-3}$ (Ref.[16,17]) and $Vol = 9.6 \times 10^{-17}$ m$^3$ (see Supporting Information Section S5). Next we performed the same experiments to extract $\delta_H$ for both DC and AC biased sensor at zero external field. Since in this case magnetic noise is dominant, we can obtain the magnetic contribution to $\delta_H$ by simply subtracting out the non-magnetic contribution ($1.3 \times 10^{-3}$) from the extracted $\delta_H$ and denoted it as $\delta_{mag}$. Figure 2b shows $\delta_{mag}$ at different bias current densities for both AC and DC bias (in the latter $j_{Pt\_rms}$ refers to the DC value). Shown in the inset is $\delta_{mag}$ at an rms current density of $5.5 \times 10^5$ A/cm$^2$ and at different bias frequencies. We can observe that $\delta_{mag}$ for DC bias (red square) is small at low current density, but increases sharply at $4 \times 10^5$ A/cm$^2$, reaches a maximum at around $6 \times 10^5$ A/cm$^2$, and finally starts to drop when the current density exceeds $7 \times 10^5$ A/cm$^2$. Such a trend is well correlated with the sensitivity dependence on current density as shown in Figure S2a, which also shows a broad maximum around $4 - 6 \times 10^5$ A/cm$^2$. This correlation can be understood from the $\delta_{mag}$ dependence on field sensitivity as derived from the FD theorem:[19]

$$\delta_{mag} = \varepsilon(f,H) \frac{2k_B T}{\pi \mu_0 M_s} \frac{\Delta R}{R} \left( \frac{1}{R} \frac{dR}{dH} \right) \quad (4)$$

where $\varepsilon(f,H)$ is the loss angle, *T* is temperature, $M_s$ is the saturation magnetization, $\mu_0$ is the vacuum permeability, $k_B$ is the Boltzmann constant, $\Delta R/R$ is the MR ratio, and $\frac{1}{R}\frac{dR}{dH}$ is the sensor's MR sensitivity. Equation (4) demonstrates clearly that when the measurement is performed at a constant temperature, $\delta_{mag}$ is proportional to the product of $\varepsilon(f,H)$ and $\frac{1}{R}\frac{dR}{dH}$. The close correlation between $\delta_{mag} \sim j_{Pt\_rms}$ and $\frac{1}{R}\frac{dR}{dH} \sim j_{Pt\_rms}$ suggests that $\varepsilon(f,H)$ is mostly independent of $j_{Pt\_rms}$ and $\delta_{mag}$ is mainly determined by $\frac{1}{R}\frac{dR}{dH}$.

However, as shown in Figure 2b, the extracted $\delta_{mag}$ exhibits a completely different trend on $j_{Pt\_rms}$ for AC bias; it decreases monotonically with $j_{Pt\_rms}$ and its value at $j_{Pt\_rms} = 5.5 \times 10^5 A/cm^2$, around 0.0027, is more than one order of magnitude smaller than the DC value of 0.04. Note that at this current density the sensitivity for both DC and AC biasing is at maximum and their values are close to each other, 1.4 mV/V/Oe for DC and 1.17 mV/V/Oe for AC bias (see Supporting Information Section S2). As shown in the inset, $\delta_{mag}$ has a negligible frequency dependence, except for the low frequency region as explained above. Therefore, the large difference in $\delta_{mag}$ between the two biasing techniques must originate from difference in the loss angle $\varepsilon(f,H)$, which is related to the energy dissipation rate. In the present case, since the biasing field is generated internally by the SOT effect, the eddy current loss can be ignored and hysteresis loss is presumably dominant. If we use the Rayleigh model to describe the hysteresis loop, the loss angle can be expressed as:[20]

$$\varepsilon = \arctan(\frac{4}{3\pi}\frac{\eta H_s}{\mu_a + \eta H_s}) \tag{5}$$

where $\eta$ is Rayleigh constant that characterizes the hysteresis, $H_s$ is the saturation field and $\mu_a$ is the initial permeability. For a soft film with large $\mu_a$ and small $H_s$ (like in the present case), the loss angle is approximately given by $\varepsilon = arctan\left(\frac{4}{3\pi}\frac{\eta H_s}{\mu_a}\right)$. The hysteresis loss is given by $\frac{4}{3}\eta H_s^3$. Therefore, a small hysteresis will lead to both a small loss angle and hysteresis loss, and thereby reducing the noise.

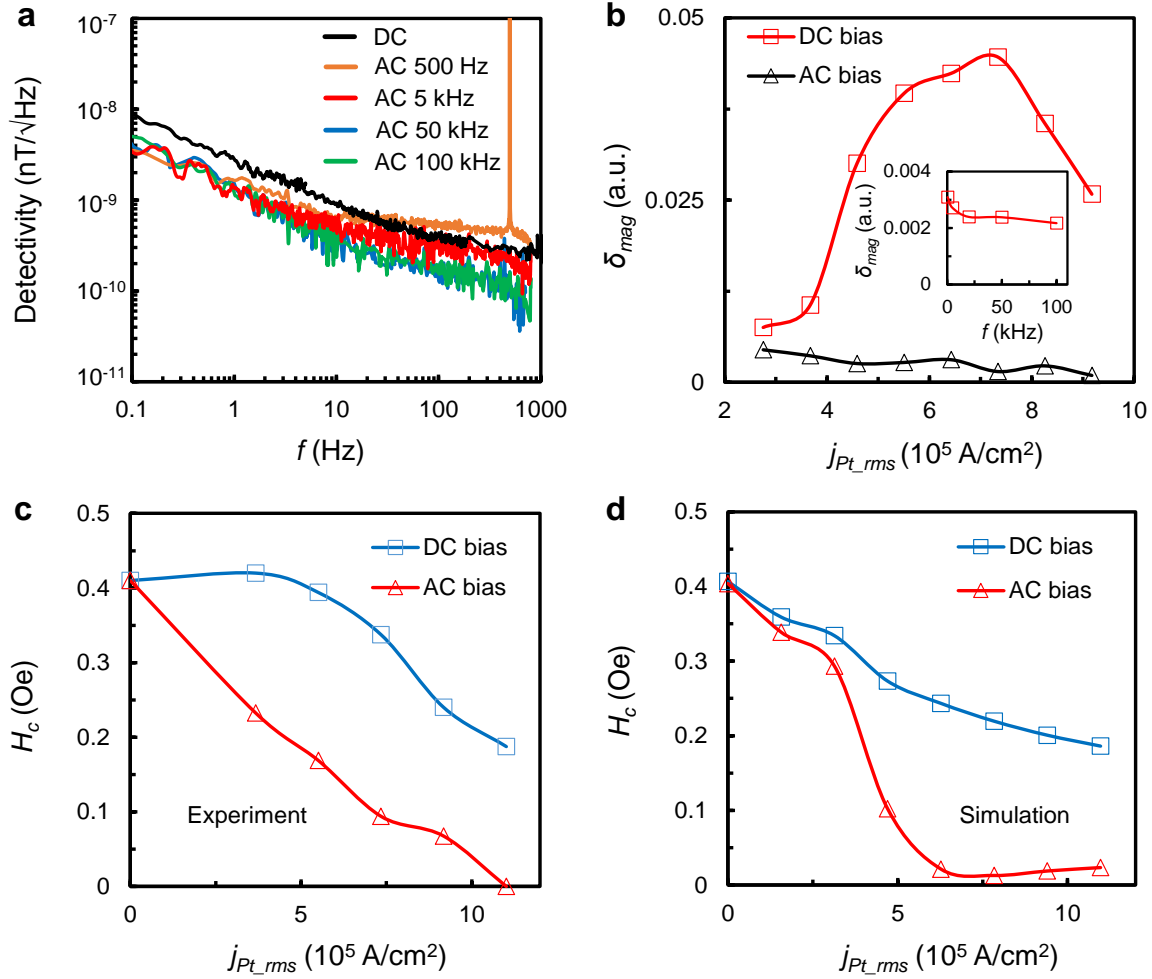

**Figure 2.** Noise and hysteresis characteristics. a) Detectivity of the SMR sensor whose field response is described in Figure 1 under DC and AC bias at different frequencies. The rms current density of the AC bias was fixed at $5.5 \times 10^5$ A/cm$^2$ for different frequencies and is the same as the DC current density. b) Extracted values of $\delta_{mag}$ at 5000 Hz with different bias current densities ($j_{Pt\_rms}$) for both AC and DC bias. The inset shows $\delta_{mag}$ at a rms current density of $5.5 \times 10^5$ A/cm$^2$ and at different frequencies. c) Comparison of the measured coercivity field ($H_c$) for a NiFe(2)/Pt(2) sensing element under DC and AC biasing as a function of current density. d) Simulated $H_c$ at different $j_{Pt\_rms}$ values using the macro-spin model. We have used the parameters: $H_k = 0.8$ Oe, $t_{Pt} = 2$ nm, and $H_{FL} = \beta\, j_{Pt}$, where $\beta = 0.51$ Oe/($10^6$ A/cm$^2$).

In order to verify if the reduction of hysteresis is indeed responsible for the noise reduction in AC biased sensors, we performed scanning magneto-optic Kerr effect (MOKE) measurements on a single sensing element under both DC and AC bias from which the hysteresis loops are extracted at different current densities. To obtain a clear hysteresis loop, the external field was swept along *x*-axis. Figure 2c compares the measured coercivity field ($H_c$) for a NiFe(2)/Pt(2) sensing element. The NiFe thickness was slightly increased from 1.8 nm to 2 nm in order to improve the signal-to-noise ratio of MOKE data. Here, $H_c$ is the average of the positive and negative coercivity fields. To exclude thermal effect as the main cause for change in hysteresis, the rms current density of AC bias was set as the same of the current density for DC bias at each measurement. As can be from the figure, the $H_c$ decreases with the increase of current density in both cases; however, the decrease in AC bias is more pronounced and it is nearly zero at $j_{Pt\_rms} = \sim 1 \times 10^6$ A/cm$^2$. Since the average power generated by the AC and DC current is the same, thermal effect can be excluded as the main cause for the much larger decrease in hysteresis in the AC bias case. Instead, the results can be understood qualitatively as caused by the presence of multiple domains inside the large sensing element. In the case of DC bias, the bias current generated an SOT effective field in *y*-direction. If the sensing element is single domain, the hysteresis will decrease quickly when the current increases based on the Stoner–Wohlfarth model. However, the decrease will not be as pronounced as in the single domain case when the sample has multiple domains, particularly when the SOT effective field is small. This may explain the results for DC biasing shown in Figure 2c. However, the situation becomes different in the case of AC biasing because in this case the SOT effective field oscillates in *y*-axis, resulting in a rotating overall field when the sweeping field in *x*-direction is small. This effectively suppresses the hysteresis due to multiple domains. To have a quantitative understanding of the difference between the two cases, we have simulated the *M-H* loop of a multi-domain ferromagnet using the macro-spin model with both a constant and time-varying bias field (Supporting Information Section S6). Figure 2d shows the simulated $H_c$ at different $j_{Pt\_rms}$ values. Although the exact shape of the two curves does not follow the experimental ones, the decreasing dependence on current density can be reproduced, and indeed $H_c$ in AC biased *M-H* loops is much smaller, in particular at high current densities. Combing the experimental and simulation results, we can conclude that the noise reduction in AC biased sensors is due to diminishing hysteresis caused by AC excitation. Therefore, we have an all-in-one magnetic sensor which features built-in excitation, diminishing hysteresis, low noise, zero offset and extremely simple structure. With all these novel features, we are ready to demonstrate a few proof-of-concept potential applications.

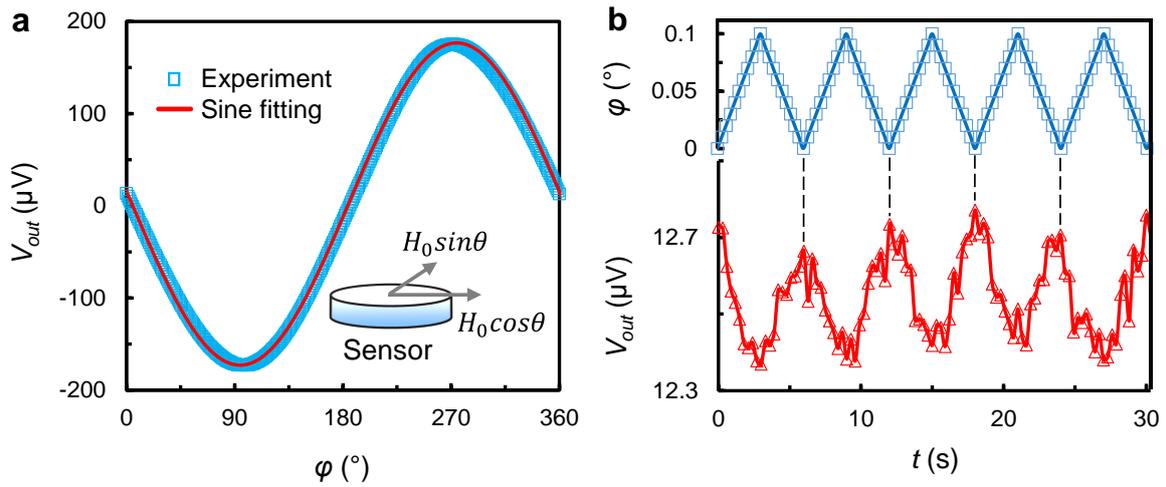

**Figure 3.** Rotation angle detection. a) Measured output voltage (blue square) as a function of $\varphi$ under $H_0 = 0.1$ Oe, together with the fitting curve to a sine function (red line). The inset shows the schematic of the rotating field generated by two in-plane field with same amplitude but a phase difference of $\pi/2$. b) Output voltage as a function of time (lower panel) when the external field angle is swept back and forth in a range of 0.1° with a step size of 0.01° (upper panel).

Angle detection is one of the major applications for magnetic sensors, including TMR, GMR and AMR sensors. For this purpose, generally a large rotating magnetic field is used to saturate the magnetization of the active layer, *i.e.*, sensing layer in AMR and free layer in TMR and GMR sensors, in the field direction. This would lead to a sinusoidal output voltage against the angle between the external field and the sensor's reference direction. The output waveform evolves one period for every 360º rotation for TMR and GMR, whereas it evolves 720º for AMR sensors (see Figure S7). Therefore, in order to measure the rotation angle, typically two TMR or GMR sensors are required, whereas in the case of the AMR, besides the two AMR sensors, a Hall sensor is also required in order to measure the rotation angle in 360º. In the case of SMR sensor, since the output voltage is proportional to $H_y$, its angle dependence should be similar to that of TMR and GMR. As the SMR sensor only has a single sensing layer without a reference, instead of saturating the magnetization in the field direction, we measured its response to a small rotating field. To this end, we placed the SMR sensor under two in-plane external field (one in $x$- and the other in $y$-direction) generated by two pairs of Helmholtz coils in a magnetically shielded cylinder. When AC currents with a same amplitude but a phase difference of $\pi/2$ are applied to the two pairs of coils, a rotating field is generated with its direction rotating continuously in the plane (see inset of **Figure 3**a). Here, $H_0$ is the amplitude of the rotating magnetic field and $\varphi$ is the angle between the rotating field and $x$-direction. By controlling the current amplitude and frequency, we can accurately control the amplitude and step-size of the rotating field. The experimentally measured output voltage as a function of $\varphi$ under $H_0 = 0.1$ Oe is shown in Figure 3a, together with the fitting curve to a sine function. It shows clearly that the output voltage from the SMR sensor exhibits an angle dependence similar to those of TMR and GMR, even though the SMR sensor has only a single magnetic layer. In order to estimate the angle resolution of the SMR sensor, we swept the field angle back and forth within the range of 0.1° and with a step size of 0.01°(upper panel of Figure 3b) and recorded the output voltage as a function of time (lower panel of Figure 3b). The clear output signal demonstrates that the SMR sensor is able to distinguish an angle difference of 0.01°, which is better than or comparable to most commercial angle sensors.[21] For actual applications, in order to suppress the influence of earth or environmental field, the sensor may be partially shielded so that the sensor will only respond to the field generated by the rotating magnet, as in most existing applications. Compared to other types of MR sensors, one has more flexibility in controlling the spacing between the sensor and the magnet considering the low-field angular sensitivity of the sensor.

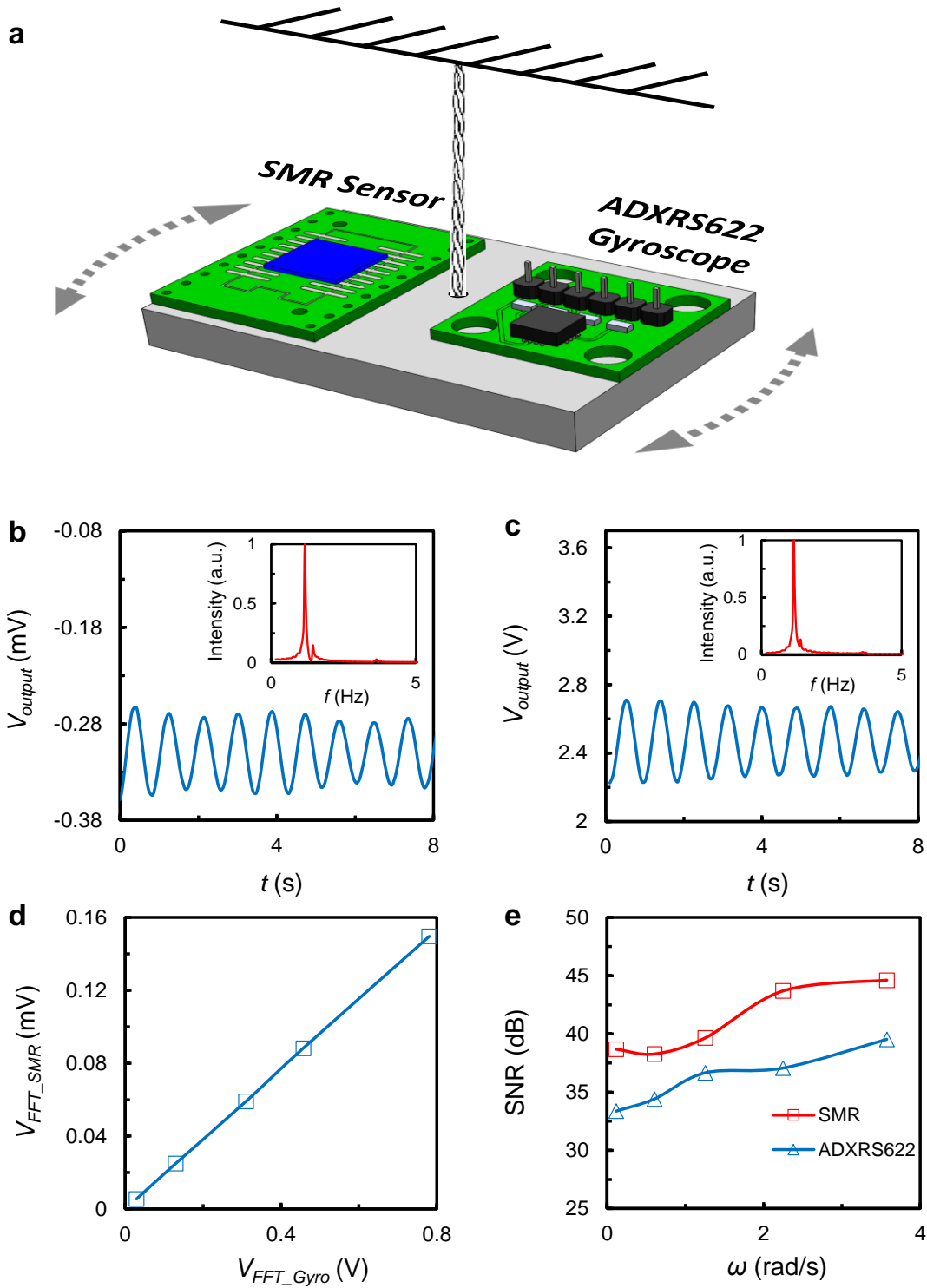

**Figure 4.** Rotational vibration detection. a) Illustration of the breadboard loaded with both a SMR sensor and an ADXRS622 gyroscope. b,c) Measured output voltage signals for the SMR sensor (b) and ADXRS622 gyroscope (c), respectively, after applying a small pulse torque. The inset shows the corresponding FFT spectrum. d) Plot of the intensity of the 1.16 Hz FFT peak of the SMR sensor against the same of ADXRS622 at different rotational amplitudes. e) Corresponding SNR as a function of the initial angular velocity for both sensors.

The high angular sensitivity of the sensor at low-field makes it promising for detection of small rotational vibration of an object. As a proof-of-concept experiment, we attached a SMR sensor consisting of a NiFe(1.8 nm)/Pt(2 nm) bilayer on a small breadboard together with a commercial gyroscope device (ADXRS622 from Analog Devices), as shown in **Figure 4**a. The electric wires for the two sensors were twisted together and fixed at pivot point to form a simple pendulum which is able to have both translational and yaw motion. The ADXRS622 comes with a preamplifier whereas the SMR sensor is just a bare Wheatstone bridge without any amplification or offset compensation; therefore, the signal levels from the two sensors are in different ranges. The measurements were performed in ambient environment with the yaw axis of the two sensors aligned in the same direction to facilitate comparison. By applying a small external torque to the pendulum, both the SMR sensor and gyroscope can detect the yaw motion. The measured output voltage signals are shown in Figure 4b and 4c for the SMR sensor and ADXRS622 gyroscope, respectively. The inset shows the corresponding fast Fourier transform (FFT) of the time-domain signals. From the FFT results, we can see that both the SMR sensor and ADXRS622 gyroscope can detect the yaw motion at 1.16 Hz. In addition, both can also detect the vibration at 1.4 Hz, presumably caused by the crosstalk from other vibration modes. To have a more quantitative comparison, we plot in Figure 4d the intensity of the 1.16 Hz FFT peak of the SMR sensor against the same of ADXRS622 measured at different vibration amplitudes. Nearly a perfect linear relation is obtained between the outputs of the two sensors. Note that the signal amplitude differs largely because the ADXRS622 comes with an amplifier whereas the SMR sensor is just a bare Wheatstone bridge without any amplification. The corresponding signal-to-noise ratio (SNR) for both sensors as a function of the initial angular velocity (calculated from the voltage output of the gyroscope and its sensitivity) is shown in Figure 4e. It is interesting to note that, despite the smaller signal amplitude, the SMR sensor exhibits a much higher SNR as compared to ADXRS622. The slight fluctuation or unevenness of the curves shown in Figure 4e is presumably due to the use of a simple experimental setup. It can certainly be improved by using a more dedicated setup for vibration studies. We have also compared the SMR sensor with commercial accelerometer in detecting vibration with both rotation and translation motions. The results are given in Supporting Information Section S7.

The combination of high sensitivity at low field and extremely simple structure makes the SMR very promising for potential applications in robotics and wearable applications. As one example, here we demonstrate the use of SMR sensor for finger motion detection. To this end, we placed a SMR sensor on the glove and then put it on the index finger, as shown in the photo of **Figure 5**a. Four wires are attached to the sensor: two for supplying AC current and the remaining two are for measuring the bridge voltage. Figure 5b illustrates the relative position of sensor detection axis with respect to the earth field when the finger is bent into different angles. Since the earth field is fixed in both strength and direction, the projection of the earth field in the sensor's detection axis direction changes with the finger bending angle, leading to different output voltages. Figure 5c shows the output voltage of one round of measurement, in which the finger is bent from vertical (90º) to 15º with respect to the earth field (See Supporting Movie 1). Clear step-wise change in the output voltage was observed, corresponding to the 6 different finger positions, as shown in the upper panel of the figure. It is worth pointing out that this is just a proof-of-concept demonstration, in which only one sensor is used. In actually applications, one may place more than one sensors on the finger, facilitating fine-motion control in robotics and virtual reality applications.

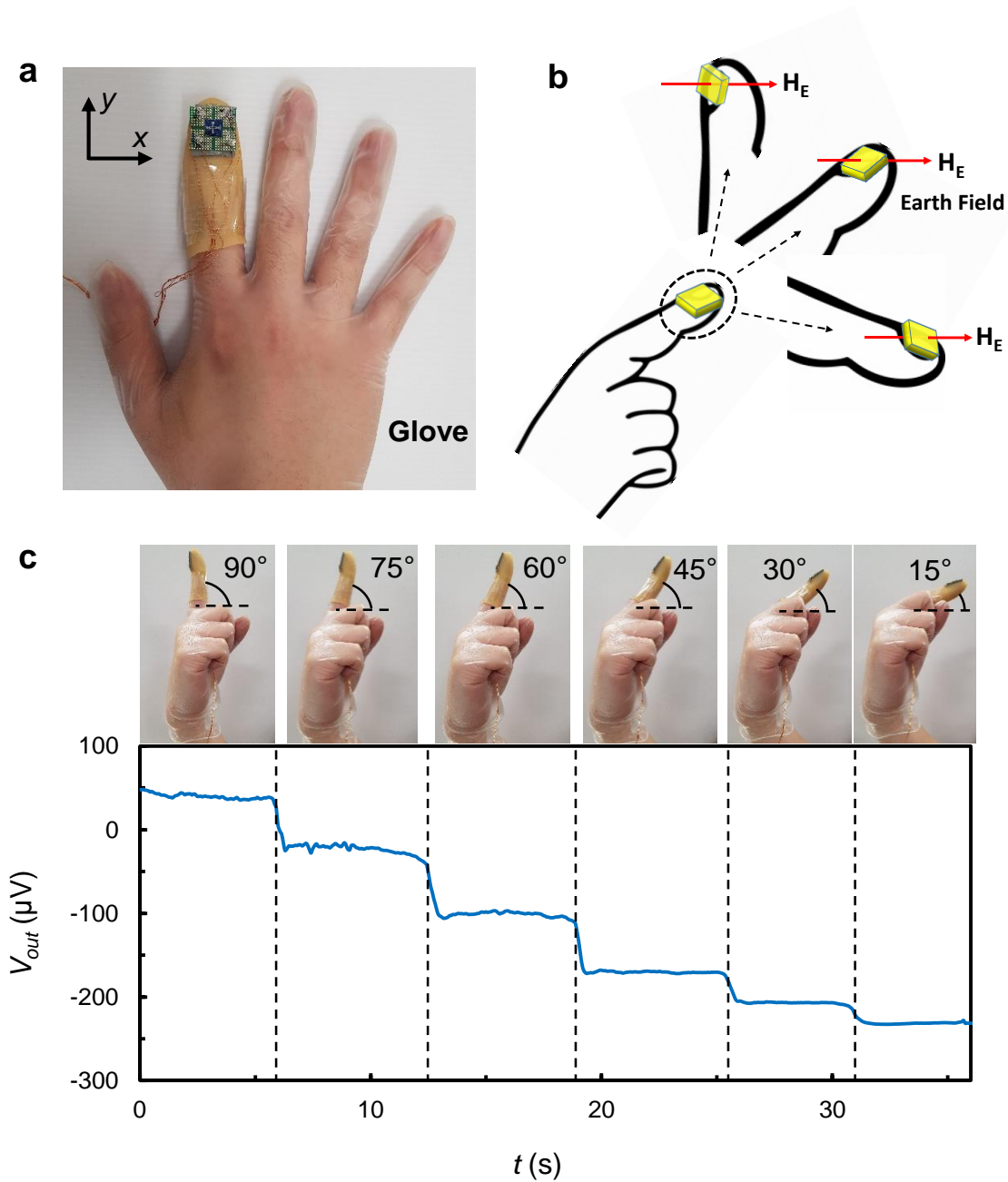

**Figure 5.** Finger motion detection. a) Hand photo showing the index finger with a SMR sensor. b) Illustration of the relative position of the sensor detection axis with respect to the earth field when the finger is bent into different angles. c) The step-wise output voltage of one round of measurement (lower panel), corresponding to the 6 different finger positions from vertical (90º) to 15º with respect to the earth field (upper panel).

In summary, we have demonstrated an all-in-one magnetic sensor that that features simplest structure, nearly zero DC offset, negligible hysteresis, and low noise by exploiting the SOT and SMR in FM/HM bilayers. The presence of SOT as a built-in bias field facilitates the implementation of AC excitation and DC detection technique, which is key to suppressing the DC offset, hysteresis and noise. A few proof-of-concept potential applications including angle, vibration, and finger movement detection have been demonstrated. This work may open new possibilities for further exploitation of the SOT technology in a variety of traditional and emerging applications.

**Experimental Section**

The sensor was fabricated on $SiO_2$/Si substrate, with the NiFe layer deposited first by evaporation followed by the deposition of Pt layer by sputtering. The base and working pressures of sputtering are $2 \times 10^{-8}$ Torr and $3 \times 10^{-3}$ Torr, respectively. Both layers were deposited in a multi-chamber system without breaking the vacuum. An in-plane field of ~500 Oe was applied during the deposition of NiFe to induce a uniaxial anisotropy in the long axis direction. The sensing elements were patterned using combined techniques of photolithography and liftoff. Before patterning into bridge sensors, thickness optimization was carried out on single sensing element and coupon films by both electrical and magnetic measurements. From these measurements, basic properties such as magnetization and magnetoresistance were obtained. Magnetic measurements were carried out using a Quantum Design vibrating sample magnetometer (VSM) with the samples cut into a size of 3 mm $\times$ 2.5 mm. The resolution of the system is better than $6\times10^{-7}$ emu. All electrical measurements were carried out at room temperature.

**Supporting Information**

Supporting Information is available from the Wiley Online Library or from the author.

**Acknowledgements**

Y.H.W. would like to acknowledge support by the Singapore National Research Foundation, Prime Minister's Office, under its Competitive Research Programme (Grant No. NRF-CRP10-2012-03). Y.H.W. is a member of the Singapore Spintronics Consortium (SG-SPIN).

**References**

[1]    a) C. Chappert, A. Fert, F. N. Van Dau, *Nat. Mater.* **2007**, 6, 813; b) S. Parkin, X. Jiang, C. Kaiser, A. Panchula, K. Roche, M. Samant, *Proc. IEEE* **2003**, 91, 661; c) Y. Wu, *Nano Spintronics for Data Storage in Encyclopedia of Nanoscience and Nanotechnology,* American Scientific Publishers, Stevenson Ranch **2003**.

[2]    a) P. Ripka, M. Janosek, IEEE Sens. J. 2010, 10, 1108; J. Lenz, S. Edelstein, *IEEE Sens. J.* **2006**, 6, 631; b) M. Díaz-Michelena, *Sens.* **2009**, 9, 2271; c) S. X. Wang, G. Li, *IEEE Trans. Magn.* **2008**, 44, 1687; d) D. L. Graham, H. A. Ferreira, P. P. Freitas, *Trends Biotechnol.* **2004**, 22, 455.

[3]    R. Bogue, *Sens. Rev.* **2014**, 34, 137.

[4]    I. M. Miron, G. Gaudin, S. Auffret, B. Rodmacq, A. Schuhl, S. Pizzini, J. Vogel, P. Gambardella, *Nat. Mater.* **2010**, 9, 230.

[5]    H. Nakayama, M. Althammer, Y. T. Chen, K. Uchida, Y. Kajiwara, D. Kikuchi, T. Ohtani, S. Geprägs, M. Opel, S. Takahashi, R. Gross, G. E. W. Bauer, S. T. B. Goennenwein, E. Saitoh, *Phys. Rev. Lett.* **2013**, 110, 206601.

[6]    Y. Yang, Y. Xu, H. Xie, B. Xu, Y. Wu, *Appl. Phys. Lett.* **2017**, 111, 032402.

[7]    Y. Xu, Y. Yang, Z. Luo, B. Xu, Y. Wu, *J. Appl. Phys.* **2017**, 122, 193904.

[8]    A. V. Silva, D. C. Leitao, J. Valadeiro, J. Amaral, P. P. Freitas, S. Cardoso, *Eur. Phys. J. Appl. Phys.* **2015**, 72, 10601.

[9]    A. Manchon, H. C. Koo, J. Nitta, S. Frolov, R. Duine, *Nat. Mater.* **2015**, 14, 871.

[10]   a) J. E. Hirsch, *Phys. Rev. Lett.* **1999**, 83, 1834; b) S. Zhang, *Phys. Rev. Lett.* **2000**, 85, 393; c) A. Hoffmann, *IEEE Trans. Magn.* **2013**, 49, 5172.

[11]   a) K. Garello, I. M. Miron, C. O. Avci, F. Freimuth, Y. Mokrousov, S. Blügel, S. Auffret, O. Boulle, G. Gaudin, P. Gambardella, *Nat. Nanotechnol.* **2013**, 8, 587; b) C. O. Avci, K. Garello, C. Nistor, S. Godey, B. Ballesteros, A. Mugarza, A. Barla, M. Valvidares,

E. Pellegrin, A. Ghosh, I. M. Miron, O. Boulle, S. Auffret, G. Gaudin, P. Gambardella, *Phys. Rev. B* **2014**, 89, 214419; c) M. Hayashi, J. Kim, M. Yamanouchi, H. Ohno, *Phys. Rev. B* **2014**, 89, 144425.

[12]   Y. Yang, Y. Xu, K. Yao, Y. Wu, *AIP Adv.* **2016**, 6, 065203.

[13]   P. Dutta, P. Horn, *Rev. Mod. Phys.* **1981**, 53, 497.

[14]   a) H. Hardner, M. Weissman, M. Salamon, S. Parkin, *Phys. Rev. B* **1993**, 48, 16156; b) S. Ingvarsson, G. Xiao, S. Parkin, W. Gallagher, G. Grinstein, R. Koch, *Phys. Rev. Lett.* **2000**, 85, 3289.

[15]   J. Almeida, R. Ferreira, P. Freitas, J. Langer, B. Ocker, W. Maass, *J. Appl. Phys.* **2006**, 99, 08B314.

[16]   M. Gijs, J. Giesbers, P. Belien, J. Van Est, J. Briaire, L. Vandamme, *J. Magn. Magn. Mater.* **1997**, 165, 360.

[17]   A. Grosz, V. Mor, E. Paperno, S. Amrusi, I. Faivinov, M. Schultz, L. Klein, *IEEE Magn. Lett.* **2013**, 4, 6500104.

[18]   Y. Guo, J. Wang, R. M. White, S. X. Wang, *Appl. Phys. Lett*. **2015,** 106, 212402.

[19]   A. Ozbay, A. Gokce, T. Flanagan, R. Stearrett, E. Nowak, C. Nordman, *Appl. Phys. Lett*. **2009**, 94, 202506.

[20]   S. Chikazumi, *Physics of magnetism*, Wiley, **1964**.

[21]   a) A. Zambrano, H. G. Kerkhoff, "Determination of the drift of the maximum angle error in AMR sensors due to aging", presented at *Mixed-Signal Testing Workshop (IMSTW), 2016 IEEE 21st International*,  **2016**; b) TDK TMR angle sensor. Available from: https://product.tdk.com/info/en/catalog/datasheets/sensor_angle-tmr-angle_en.pdf